\documentclass[aps,prl,reprint,showpacs]{revtex4-1}
\usepackage{amsmath}
\usepackage{graphicx}
\usepackage{epstopdf}
\usepackage{bbold}
\usepackage{color}
\usepackage{float}
\usepackage[caption = false]{subfig}

\begin{document}
\title{Minimal Entanglement Witness From Electrical Current Correlations}
\author{F. Brange}
\author{O. Malkoc}
\author{P. Samuelsson}
\affiliation{Department of Physics, Lund University, Box 118, SE-221 00 Lund, Sweden}


\begin{abstract}
Despite great efforts, an unambiguous demonstration of entanglement of mobile electrons in solid state conductors is still lacking. Investigating theoretically a generic entangler-detector setup, we here show that a witness of entanglement between two flying electron qubits can be constructed from only two current cross correlation measurements, for any nonzero detector efficiencies and non-collinear polarization vectors. We find that all entangled pure states, but not all mixed ones, can be detected with only two measurements, except the maximally entangled states, which require three. Moreover, detector settings for optimal entanglement witnessing are presented.
\end{abstract}

\maketitle

\emph{Introduction. } Over the last two decades, the demand for efficient methods to detect entanglement, key to quantum information processing \cite{Nielsen&ChuangI}, has spurred research on entanglement witnesses \cite{Horodecki19961,Terhal2000319,PhysRevA.62.052310,Gühne20091,RevModPhys.81.865}. A witness is an observable quantity which, for at least one entangled state, takes on a value outside the range accessible for separable (non-entangled) states. Entanglement detection with witnesses has been demonstrated with e.g. photons \cite{PhysRevLett.91.227901, PhysRevLett.92.087902, PhysRevA.82.012302}, ions \cite{Roos1478,Nature2005,PhysRevLett.106.130506} and atomic nuclei \cite{Filgueiras2012}. Importantly, for any entangled state, including locally realistic ones \cite{PhysRevA.40.4277}, there is a witness that can detect it \cite{Horodecki19961}. Moreover, witnesses, in contrast to Bell inequalities \cite{Bell1964,PhysRevLett.23.880} and full tomographic reconstructions \cite{PhysRevA.64.052312,PhysRevA.66.012303}, allow entanglement detection with only a few local measurements \cite{PhysRevA.66.062305, GühneToth2006}, even when the number of entangled particles is large \cite{PhysRevLett.94.060501}.

The prospects of few-measurement entanglement detection make witnesses particularly interesting for flying qubits in solid state conductors, where an unambiguous demonstration of entanglement is still lacking. Here, detection schemes for spatially separated, spin \cite{doi:10.1143/JPSJ.70.1210,PhysRevB.66.161320} or orbitally \cite{PhysRevLett.91.157002,PhysRevLett.91.147901,PhysRevLett.92.026805,PhysRevB.73.041305} entangled, electrons have been proposed based on experimentally accessible current cross correlations \cite{Blanter20001}. However, the required set of measurements, with different, non-collinear detector settings and correlations between two or more pairs of detector terminals, are experimentally highly challenging. Aiming for less demanding measurements, works \cite{PhysRevB.75.165327, PhysRevB.89.125404} on witnesses have proposed schemes with only two or three settings and less than ten cross correlations, allowing detection of certain classes of entangled states. Yet, two fundamental questions remain unanswered: (i)~What is the minimum number of current cross correlation measurements needed for an entanglement witness? (ii)~Which entangled states can be detected by such a witness?

\begin{figure}[h]
    \centering
    \includegraphics[width=0.45\textwidth]{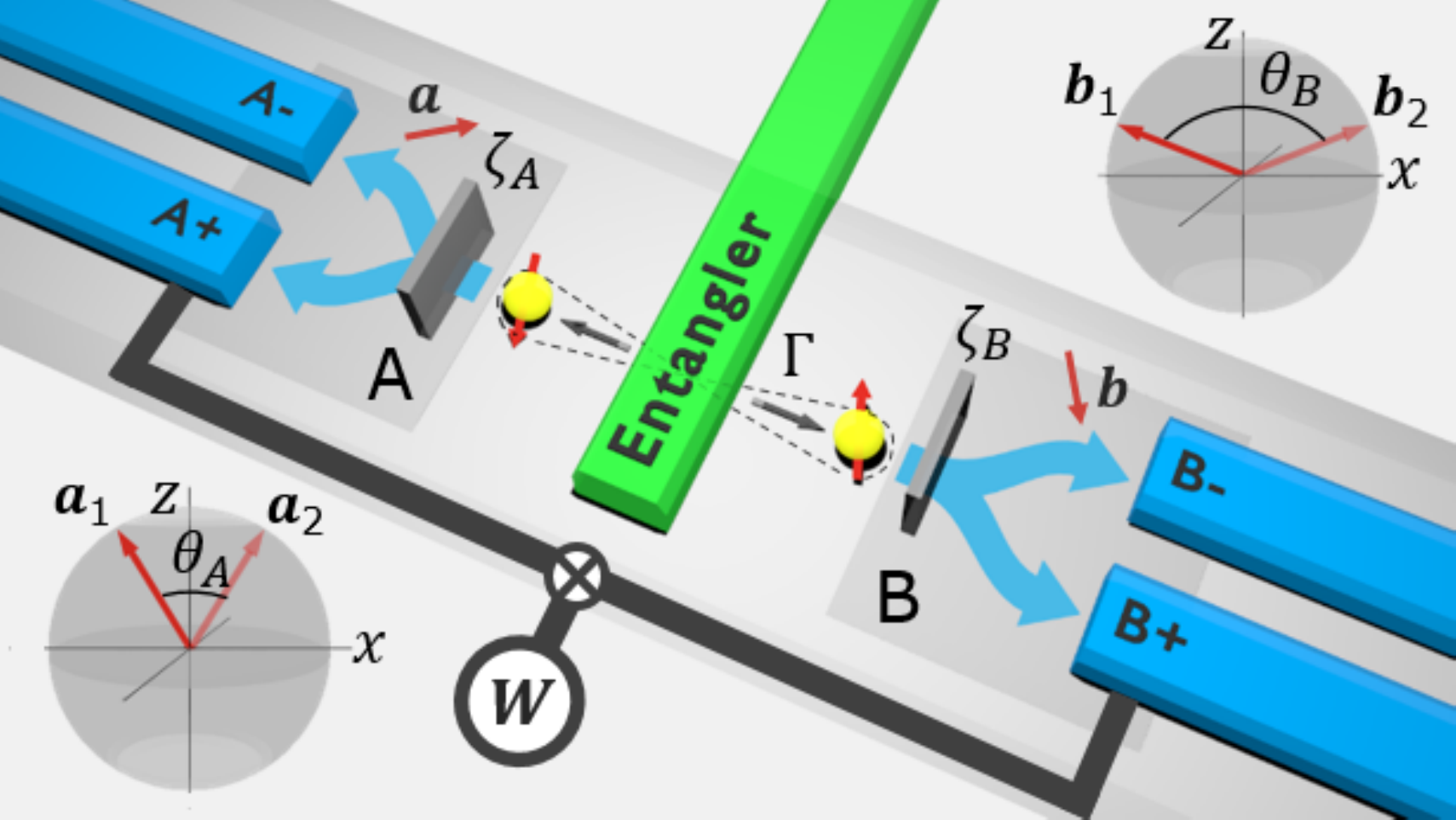}
    \captionsetup{justification=justified,singlelinecheck=false}
    \caption{Schematic of the generic entangler-detector setup, consisting of an entangler and two detector systems $A$ and $B$. The entangler generates split pairs of entangled electrons, in either spin or orbital degrees of freedom, at a rate $\Gamma$. Each detector system consists of two detector terminals ($\pm$) and a beam splitter with efficiency $0\leq \zeta_A, \zeta_B \leq 1$ and unit-magnitude polarization vector $\textbf{a}, \textbf{b}$. An entanglement witness $W$ can be constructed from only two current cross correlation measurements, with respective polarization vectors $\textbf{a}_1, \textbf{a}_2$ at $A$ ($\textbf{b}_1, \textbf{b}_2$ at $B$). The local coordinate systems are defined such that $\textbf{a}_1$ and $\textbf{a}_2$ ($\textbf{b}_1$ and $\textbf{b}_2$) lie in the $xz$-plane, symmetrically about the $z$-axis, with relative local angles $\theta_A$ ($\theta_B$). The correlations can be measured between currents at two terminals only (here $A+$ and $B+$).}
    \label{Entangler-detector}
\end{figure}

Here we answer these two questions within a generic solid state entangler-detector model, see Fig.~\ref{Entangler-detector}. We find that only two cross correlation measurements -- two detector settings with one measurement per setting -- are sufficient to constitute a witness. Moreover, we show that all entangled pure (but not all mixed) states can be detected by the witness, except the maximally entangled, which require three measurements. In addition, the most favorable settings for detecting a given pure state are identified. Our findings will greatly simplify detection of entanglement between mobile electrons in solid state conductors, such as the recently investigated Cooper pair splitter \cite{Nature.461.960, PhysRevLett.104.026801,PhysRevLett.107.077005, Nat.Comm.3.1165, PhysRevLett.109.157002}.

\emph{Entangler-detector model. } A generic entangler-detector setup is shown in Fig.~\ref{Entangler-detector}. The entangler emits, at a rate $\Gamma$, identically prepared split pairs of entangled electrons towards the detector systems $A$ and $B$. The electrons are either spin \cite{PhysRevB.63.165314,PhysRevB.65.165327,PhysRevLett.89.037901} or orbitally \cite{PhysRevLett.84.5912,PhysRevLett.91.147901, PhysRevLett.91.157002, PhysRevLett.93.056803, PhysRevB.73.235331,PhysRevLett.92.026805,Nature.448.333, PhysRevLett.103.056802, PhysRevLett.114.176803} entangled, with local basis states denoted $|\!\!\uparrow\rangle$, $|\!\! \downarrow\rangle$, hereafter referred to as spin up/down. Detector system $A$ ($B$) consists of a polarizing beam splitter, characterized by a unit polarization vector \textbf{a} (\textbf{b}), and two detector terminals $A\pm$ ($B\pm$). All electrons arriving at the detector terminals contribute to the electric current, i.e., the terminals have ideal efficiencies. To account for e.g. dephasing and/or non-perfect polarization of ferromagnetic detector terminals \cite{PhysRevB.89.125404,Malkoc2014}, the beam splitter efficiencies are $0\leq \zeta_A, \zeta_B\leq 1$. An electron incident on the beam splitter at $A$ will consequently, with a probability $\zeta_A$, be projected along the spin direction defined by $\mathbf{a}$ and transmitted to the respective detector $A\pm$. With a probability $1-\zeta_A$ the particle is instead, irrespective of its spin, transmitted with equal probability to $A+$ and $A-$. The same holds for detector system $B$.

For flying electron qubits, real-time detection of quantum properties of individual particles is presently out of reach. Instead, the natural candidate for entanglement detection is low-frequency cross correlations of electric currents flowing into detector terminals $A\pm$ and $B\pm$. Focusing on an entangler-detector system operating in the tunneling regime, we can write the current cross correlators as $S^{\pm \pm}_{AB}(\textbf{a},\textbf{b}) = e^2 \Gamma \text{tr}[(\mathbb{1}\pm \zeta_A\mathbf{a}\cdot\boldsymbol{{\sigma}})\otimes(\mathbb{1}\pm \zeta_B\mathbf{b}\cdot\boldsymbol{{\sigma}})\rho]$. Here $e$ is the elementary charge, $\boldsymbol{\sigma} = (\sigma_x,\sigma_y,\sigma_z)$ is a vector of Pauli matrices, $\otimes$ is the tensor product and $\rho$ is the two-particle spin density matrix of the pair emitted by the entangler.

\emph{Witness. } For an entanglement witness based on current cross correlations to be experimentally viable, it is key to minimize the number of correlations measurements as well as the number of detector terminal pairs $A\pm,B\pm$ between which currents need to be correlated. In addition, it is important to minimize the number of distinct detector settings and to find a set of polarization vectors, $\textbf{a}_i$ at $A$ and $\textbf{b}_i$ at $B$, in the same plane. Motivated by these requirements we seek a witness based on measurements of $N$ cross correlations between currents at two detector terminals only. Taking, without loss of generality, terminals $A+$ and $B+$ the corresponding linear operator is
\begin{equation}
W^{(N)} = \sum_{i=1}^{N}(\mathbb{1}+\zeta_A\textbf{a}_i\cdot\boldsymbol{{\sigma}})\otimes(\mathbb{1}+\zeta_B\textbf{b}_i\cdot\boldsymbol{{\sigma}}).
\label{Witness form}
\end{equation}
For $W^{(N)}$ to be an entanglement witness there needs to be \cite{Horodecki19961,Terhal2000319} at least one entangled state $\rho$ such that one, upper $(+)$ or lower $(-)$, detector marginal $\Delta^{W\pm}_\rho$ satisfies
\begin{eqnarray}
\nonumber
 \Delta^{W+}_\rho &\equiv&\text{tr}(W^{(N)}\rho)-\max \limits_{\rho_s}\text{tr}(W^{(N)}\rho_s) > 0,\\
 \Delta^{W-}_\rho &\equiv& \min \limits_{\rho_s}\text{tr}(W^{(N)}\rho_s)-\text{tr}(W^{(N)}\rho) > 0,
\label{Witness condition}
\end{eqnarray}
where the maxi-/minimization is carried out over all separable states $\rho_s$. We stress that to experimentally test the entanglement of $\rho$ via Eq.~\eqref{Witness condition}, the quantity $\text{tr}(W^{(N)}\rho) = (e^2 \Gamma)^{-1} \sum_{i=1}^NS^{++}_{AB}(\textbf{a}_i,\textbf{b}_i)$ obtained from the cross correlation measurements ($\Gamma$ obtained from independent average current measurements) should be compared to the calculated value $\text{max/min}_{\rho_s}\text{tr}(W^{(N)}\rho_s)$. Our main questions can now be rephrased as: (i) What is the minimum $N$ required to make $W^{(N)}$ fulfill Eq.~\eqref{Witness condition}? (ii) Which entangled states can be detected by this $W^{(N)}$?

\emph{Key results. } We first summarize our answers to these two questions; more details follow below. (i) While $W^{(1)}$ is a tensor product of local operators at $A$ and $B$, and hence no witness, already $W^{(2)}$ is found to be a witness for all detector parameters $\zeta_A$, $\zeta_B$, $\textbf{a}_i$, $\textbf{b}_i$, except a set of special cases. For a bipartite system of qubits, $W^{(2)}$ is a witness if and only if the eigenstate corresponding to either the smallest or the largest eigenvalue is (a) non-degenerate and (b) entangled \cite{Wallach}. To show (a), we evaluate the eigenvalues $\lambda_i$ of $W^{(2)}$ giving
\begin{equation}
\lambda_i = 2\left(1\pm c_Ac_B\pm\sqrt{(c_A\pm c_B)^2+s_A^2s_B^2} \right),
\label{eigenvalues}
\end{equation}
where the indices $i=1,2,3,4$ correspond to an order of $\pm$ as $\{+-+\}$,$\{---\}$,$\{-+-\}$,$\{+++\}$, with $\lambda_1\leq\lambda_2\leq\lambda_3\leq\lambda_4$, depending only on $c_\alpha = \zeta_\alpha \cos(\theta_\alpha/2)$, $s_\alpha = \zeta_\alpha \sin(\theta_\alpha/2)$, $\alpha = A, B$, with $\cos(\theta_A) = \textbf{a}_1\cdot \textbf{a}_2$ and $\cos(\theta_B) = \textbf{b}_1\cdot \textbf{b}_2$. Both eigenvalues $\lambda_1, \lambda_4$ are non-degenerate for all detector parameters, except $\theta_A = \pi$, $\theta_B = \pi$, $\zeta_A = 0$ or $\zeta_B=0$, for which $\lambda_1 = \lambda_2$ and $\lambda_3 = \lambda_4$ due to local rotation symmetries of the polarization vectors, and $\zeta_A=\zeta_B=1$ for which $\lambda_1=\lambda_2$ due to a hidden, non-local symmetry \cite{0305-4470-36-10-317,Hou&Chen}.

To show (b), we introduce, without loss of generality, local coordinate systems as shown in Fig.~\eqref{Entangler-detector}, yielding the eigenstates $|\psi_1\rangle$, $|\psi_4\rangle$ corresponding to $\lambda_1$, $\lambda_4$ as
\begin{eqnarray}
\nonumber
|\psi_1\rangle&=&\sin\alpha|\! \uparrow \uparrow\rangle - \cos\alpha|\! \downarrow \downarrow\rangle, \\
|\psi_4\rangle&=&\cos \alpha|\! \uparrow \uparrow\rangle + \sin \alpha|\! \downarrow \downarrow\rangle, 
\label{smallest eigenstate}
\end{eqnarray}
 in the local $\{ |\!\!\uparrow \uparrow\rangle, |\!\!\uparrow \downarrow\rangle, |\!\!\downarrow \uparrow\rangle, |\!\!\downarrow \downarrow\rangle \}$-basis with the angle $0\leq\alpha\leq\pi/4$ given by
\begin{equation}
\tan \alpha = \Big(\sqrt{(c_A+c_B)^2+s_{A}^2s_B^2}-(c_A+c_B)\Big)/s_{A}s_B.
\label{alpha}
\end{equation}
The states $|\psi_1\rangle$, $|\psi_4\rangle$ are entangled for all $\alpha \neq 0$, i.e., for all detector parameters except $\theta_A=0$, $\theta_B=0$, $\zeta_A=0$ or $\zeta_B=0$, for which $W^{(2)}$ can be written as a product of local operators. \emph{Hence, the answer to (i) is $N=2$; $W^{(2)}$ is a witness operator for all non-zero detector efficiencies ($\zeta_A, \zeta_B >0$) and non-collinear polarization vectors ($\theta_A, \theta_B \neq 0, \pi$).}

\begin{figure}[t]
    \centering
    \includegraphics[width=0.45\textwidth]{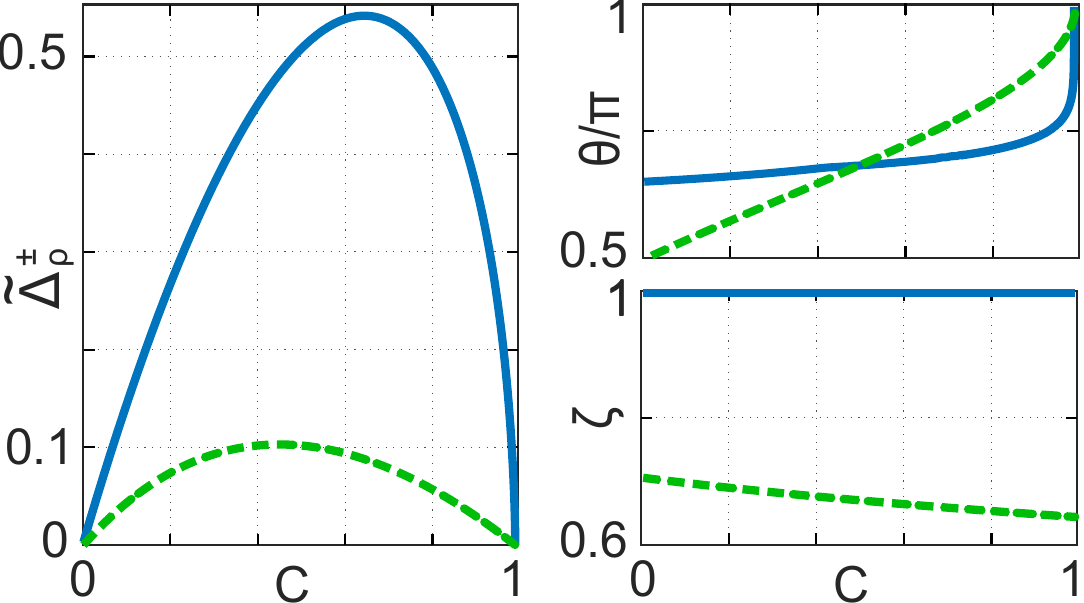}
    \captionsetup{justification=justified,singlelinecheck=false}
    \caption{Left: The maximal detection marginals $\tilde{\Delta}_{\rho}^{\pm}$ of $W^{(2)}$ for pure states $\rho$ with concurrence $0\leq C \leq 1$. Right: Symmetric detector settings $\theta_A=\theta_B=\theta$ (top) and $\zeta_A=\zeta_B=\zeta$ (bottom) giving $\tilde{\Delta}_{\rho}^{\pm}$ of $W^{(2)}$ in left panel. In all plots, solid blue (dashed green) line corresponds to $\tilde{\Delta}_{\rho}^{+}$ ($\tilde{\Delta}_{\rho}^{-}).$}
    \label{Delta sym}
\end{figure}

\begin{figure}[t]
    \centering
    \includegraphics[width=0.44\textwidth]{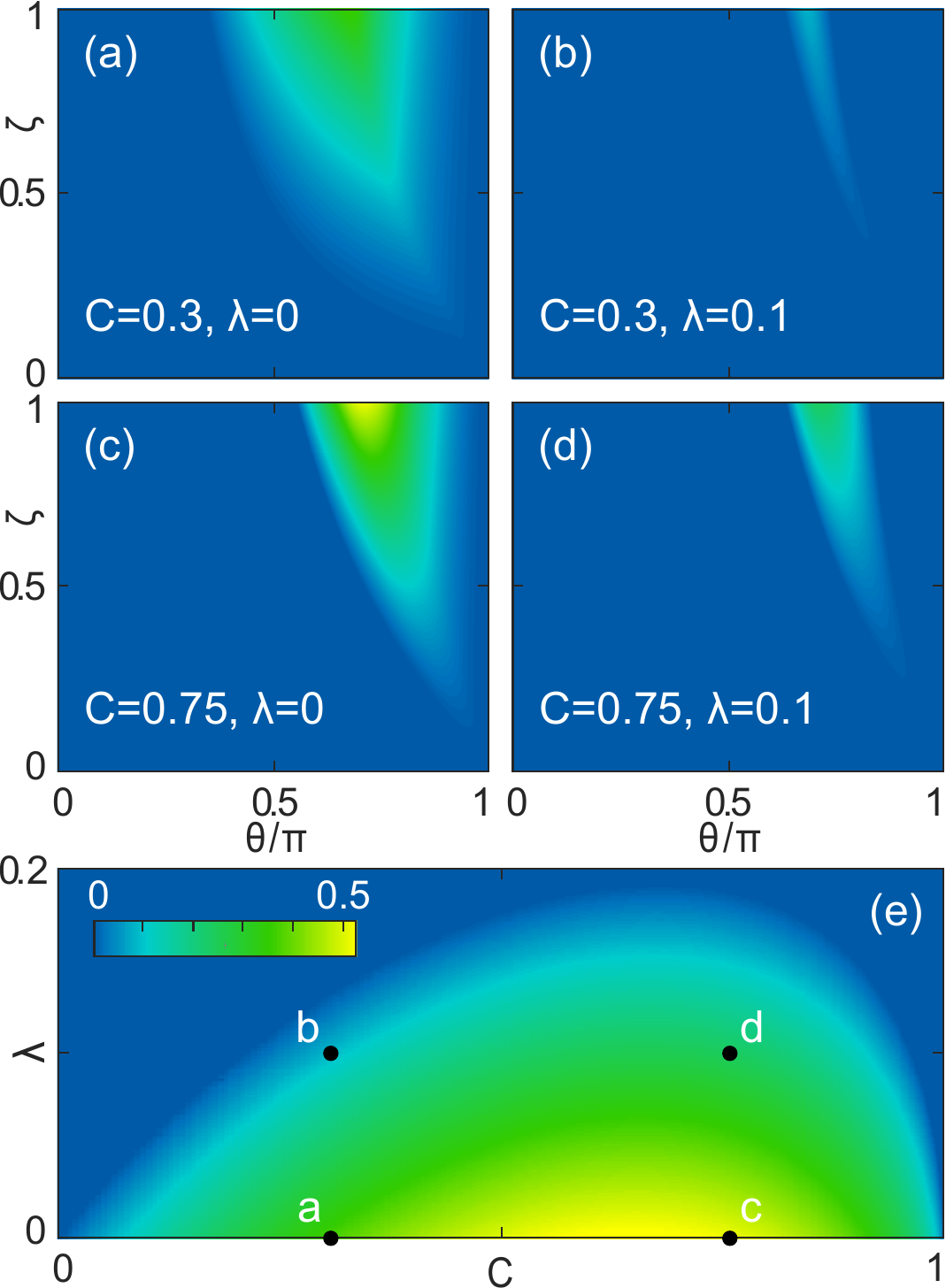}
    \captionsetup{justification=justified,singlelinecheck=false}
    \caption{(a)--(d) The detection marginal $\Delta^{W+}_\rho$ for symmetric setups, $\theta_A=\theta_B=\theta$ and $\zeta_A=\zeta_B=\zeta$, with mixed states, $\rho = (1-\lambda)|\psi\rangle\langle\psi|+\lambda \mathbb{1}/4$, $0\leq\lambda\leq 1$, where $|\psi\rangle$ has concurrence $0\leq C\leq 1$. (e) $\tilde{\Delta}^{+}_\rho$ for $\rho$ with different $C$ and $\lambda$. The cases in (a)--(d) are marked with black dots. The color scale in (e) holds also for panels (a)--(d).}
    \label{Delta for different witnesses}
\end{figure}

To answer (ii), for a given entangled $\rho$ we calculate $\tilde{\Delta}^\pm_\rho$, the maximum detector marginal $\Delta^{W\pm}_\rho$ for any $W^{(2)}$; $\tilde{\Delta}^\pm_\rho>0$ ($\tilde{\Delta}^\pm_\rho\leq 0$) implies that $\rho$ can (can not) be detected by at least one (any) $W^{(2)}$. For a pure state, where $\tilde{\Delta}^\pm_\rho$ is a unique function of the entanglement of $\rho$, we find (see Fig.~\ref{Delta sym}) that $\tilde{\Delta}^+_\rho>\tilde{\Delta}^-_\rho>0$ for all non-maximally entangled $\rho$, while $\tilde{\Delta}^+_\rho=\tilde{\Delta}^-_\rho=0$ for the maximally entangled ones. This result does not in general extend to mixed states, as illustrated in Fig.~\ref{Delta for different witnesses}. \emph{Hence, the answer to (ii) is: all entangled pure (but not all mixed) states, except the maximally entangled ones, can be detected by at least one $W^{(2)}$}. Symmetric parameter settings, maximizing the detector marginal $\Delta^{W\pm}_\rho$, are shown in Fig.~\ref{Delta sym}. By performing a third measurement, i.e. for $W^{(3)}$, the maximally entangled states can be detected.

\emph{Derivation of results.}  To arrive at our results we start by recalling \cite{Terhal2000319} some basic properties of entanglement witnesses. The spectral decomposition of the operator in Eq. (\ref{Witness form}) is $W^{(N)}=\sum_i \lambda_i^{(N)}|\psi_i^{(N)}\rangle \langle \psi_i^{(N)}|$, with eigenvalues ordered as 
$\lambda_1^{(N)} \leq \lambda_2^{(N)}\leq \lambda_3^{(N)}\leq \lambda_4^{(N)}$. Considering $\Delta_{\rho}^{W-}$ in Eq. (\ref{Witness condition}), by choosing the state  $\rho=|\psi_1^{(N)}\rangle \langle \psi_1^{(N)}|$ the term $\mbox{tr}(W^{(N)}\rho)=\lambda_1^{(N)}$, taking on its smallest possible value. For $|\psi_1^{(N)}\rangle$ entangled, any separable state $\rho_s$ written in the eigenbasis of $W^{(N)}$ will have a non-zero weight of at least one state $|\psi_i^{(N)}\rangle$ with $i \geq 2$. This gives $\min_{\rho_s}\text{tr}(W^{(N)}\rho_s) \geq \lambda_1^{(N)}$ and hence $\Delta^{W-}_\rho \geq 0$. Similar arguments, based on $\lambda_4^{(N)}$ and $|\psi_4^{(N)}\rangle$, hold for the upper marginal $\Delta^{W+}_\rho$.
 
Importantly, to guarantee strictly $\Delta^{W\pm}_\rho > 0$, for a two-qubit system the entangled eigenstate $|\psi_{1}^{(N)}\rangle$ or $|\psi_{4}^{(N)}\rangle$ also needs to be non-degenerate; for a two-fold or higher degeneracy there exists at least one separable linear combination of degenerate eigenstates \cite{Wallach}, resulting in $\Delta^{W\pm}_\rho = 0$. Hence, to answer (i), we need to find the minimum $N$ for which $W^{(N)}$ has an eigenstate $|\psi_{1}^{(N)}\rangle$ or $|\psi_{4}^{(N)}\rangle$  which is both (a) unique and (b) entangled.

Starting at $N=1$, the eigenstates of $W^{(1)}$, a tensor product of local operators, are all separable, and hence $W^{(1)}$ is no witness. For $N=2$, to test (a), we calculate the eigenvalues of $W^{(2)}$, given in Eq.~\eqref{eigenvalues}, defining $\lambda_i\equiv\lambda_i^{(2)}$. It is clear from Eq.~\eqref{eigenvalues} that for all detector efficiencies $\zeta_A$, $\zeta_B$ and relative, local angles $\theta_A$, $\theta_B$, we have $\lambda_1 \leq \lambda_2 \leq \lambda_3 \leq \lambda_4$. Only in three distinct cases we have degeneracies: for 1) anti-parallel polarization vectors, $\theta_A=\pi$ or $\theta_B = \pi$, and 2) zero efficiency, $\zeta_A=0$ or $\zeta_B = 0$, both the largest and the smallest eigenvalue are doubly degenerate, $\lambda_1=\lambda_2$ and $\lambda_3 = \lambda_4$ while  for 3) ideal detector efficiencies, $\zeta_A=\zeta_B=1$, only the smallest eigenvalue is degenerate, $\lambda_1=\lambda_2$ but $\lambda_3\neq\lambda_4$. 

The degeneracies are consequences of underlying symmetries of the witness operator. In case 1), with polarization vectors at detector system $A$ ($B$) obeying $\textbf{a}_1=-\textbf{a}_2\equiv \textbf{a}$ ($\textbf{b}_1=-\textbf{b}_2\equiv \textbf{b}$), $W^{(2)}$ is invariant under local rotations $U_\textbf{a} = \exp(-i\frac{\varphi}{2}\textbf{a}\cdot\boldsymbol{\sigma})\otimes \mathbb{1}$ ($U_\textbf{b} = \mathbb{1}\otimes \exp(-i\frac{\varphi}{2}\textbf{b}\cdot\boldsymbol{\sigma})$) with $\varphi$ an arbitrary angle. For case 2) $W^{(2)}$ is independent of the polarization vectors at detector system $A$ ($B$) and invariant under any local rotation of them. In case 3) the degeneracy arises because of an underlying non-local symmetry \cite{0305-4470-36-10-317,Hou&Chen}, with no transparent physical origin.

To test condition (b) for $W^{(2)}$, we first note that a defining property of entanglement is that it is invariant under local unitary operations, i.e., independent of the choice of local coordinate systems. We can thus, conveniently, take the polarization vectors symmetrical about the $z$-axis in the $xz$-plane, as illustrated in Fig.~\ref{Entangler-detector}, giving
\begin{equation}
W^{(2)} = 2[s_As_B\sigma_x\otimes\sigma_x+(\mathbb{1}+c_A\sigma_z)\otimes(\mathbb{1}+c_B\sigma_z)],
\label{Witness simpl}
\end{equation}
depending, as $\lambda_i$, only on the detector parameters $\theta_A$, $\theta_B$, $\zeta_A$, $\zeta_B$ via $c_{A/B}$ and $s_{A/B}$. In the same basis, the eigenstates $|\psi^{(2)}_1\rangle \equiv |\psi_1\rangle$ and $|\psi^{(2)}_4\rangle \equiv |\psi_4\rangle$ corresponding to the smallest, $\lambda_1$, and largest, $\lambda_4$, eigenvalue respectively, are given by Eq.~\eqref{smallest eigenstate}, with Schmidt coefficients \cite{Schmidt} $\cos \alpha$ and $\sin \alpha$, where $\alpha$ is defined by Eq.~\eqref{alpha}. The entanglement can be quantified in terms of the concurrence $C$ \cite{PhysRevLett.80.2245} ranging from zero for a separable state to unity for a maximally entangled state. For both $|\psi_1\rangle$ and $|\psi_4\rangle$ in Eq.~\eqref{smallest eigenstate}, we have
\begin{equation}
C=\sin(2\alpha), \quad 0\leq \alpha \leq \pi/4,
\end{equation}
i.e., $|\psi_1\rangle$ and $|\psi_4\rangle$ are separable only for $\theta_A,\theta_B=0$ or $\zeta_A,\zeta_B=0$, for which $W^{(2)}$ is a tensor product of local operators [clear from Eq.~\eqref{Witness simpl}]. For all other detector parameters, $|\psi_1\rangle$ and $|\psi_4\rangle$ are entangled and condition (b) is fulfilled. \emph{We thus see that $W^{(2)}$ is a witness operator for all detector parameters, except $\theta_A,\theta_B = 0,\pi$, $\zeta_A,\zeta_B = 0$ or $\zeta_A=\zeta_B=1$, answering question (i).}

To answer (ii), we first note that a sufficient, but not necessary, condition for a state to be detectable is that it can be written as in Eq.~\eqref{smallest eigenstate} for an $\alpha$ that is given via Eq.~\eqref{alpha} for at least one witness $W^{(2)}$. This condition is fulfilled for all non-maximally entangled states, although the corresponding $W^{(2)}$ is not necessarily the most optimal choice for detection. To find the optimal witness and to give a necessary condition, we consider the quantity $\tilde{\Delta}^\pm_\rho \equiv \max_{W^{(2)}} \Delta^{W\pm}_\rho$, the largest possible detection marginal. Importantly, for $\tilde{\Delta}^\pm_\rho>0$ ($\tilde{\Delta}^\pm_\rho\leq 0$) the state $\rho$ can (can not) be detected for at least one (any) $W^{(2)}$. The calculation proceeds in two steps. First, the maxi-/minimization over all separable states $\rho_s$ in Eq.~\eqref{Witness condition}, respectively, is carried out \footnote{It suffices to maximize over the pure separable states \unexpanded{$\rho_s = |\psi_s\rangle\langle \psi_s|$} parametrized as \unexpanded{$|\psi_s\rangle = |\psi_A\rangle \otimes |\psi_B\rangle$}, \unexpanded{$|\psi_{\alpha}\rangle=\cos \phi_{\alpha}|\!\uparrow \rangle+e^{i\varphi_{\alpha}}\sin \phi_{\alpha}|\!\downarrow \rangle$}, \unexpanded{$\alpha = A,B$}}, giving

\begin{eqnarray}
 \underset{\rho_s}{\text{max/min }}\mbox{tr}(W^{(2)}\rho_s)&=&\left\{
                \begin{array}{ll}
                  X^{\pm}_1,  \quad W^{(2)}\in R_1\\
                  X^{\pm}_2,  \quad W^{(2)}\in R_2
                \end{array}
              \right.
\end{eqnarray}
where $+$ ($-$) corresponds to the maximized (minimized) value and the boundary between parameter regions $R_1$ and $R_2$ is defined by $X^\pm_1 = X^\pm_2$. Here $X^\pm_1 = 2[1\pm c_A \cos(\theta_A/2)][1\pm c_B\cos(\theta_B/2)]$ and $X^\pm_2 = \frac{ \mp\sqrt{Y_{AB}Y_{BA}}-(1-\cos \theta_A)(1-\cos \theta_B)}{\cos\theta_A+\cos\theta_B}$, with $Y_{AB} = (1-\cos \theta_A)(1+\cos \theta_A-\zeta_B^2(\cos \theta_A+\cos \theta_B)$ and $Y_{BA}$ is obtained from $Y_{AB}$ by interchanging all subscripts $A\leftrightarrow B$.

Second, we perform the maximization over the parameters of $W^{(2)}$ numerically. Importantly, a pure two-qubit state is, up to a local unitary operation, uniquely characterized by its entanglement \cite{LindenPopescu1998,PhysRevA.60.910,Vidal2000}. Since $\tilde{\Delta}_\rho ^{\pm}$ is invariant under unitary operations of $\rho$, it is hence a function of the entanglement of $\rho$ only. The obtained $\tilde{\Delta}_\rho ^{\pm}$ for entangled pure states $\rho$, characterized by $C>0$, is shown in Fig.~\ref{Delta sym}; we have $\tilde{\Delta}^{+}_\rho > \tilde{\Delta}^{-}_\rho>0$, with the only exception being states with $C=1$ for which $\tilde{\Delta}^{+}_\rho = \tilde{\Delta}^{-}_\rho=0$, see discussion below. \emph{In other words, all entangled pure states, except the maximally entangled, are detectable by $W^{(2)}$, answering question (ii).}

This result naturally raises the directly related question: How to choose the detector settings to have the best possible witness detection of a given entangled pure state? That is, what parameters of $W^{(2)}$ maximize $\Delta^{W\pm}_\rho$ for a $\rho=|\psi\rangle\langle \psi|$? We identify \footnote{See Supplemental Material at ... for further details.} one particularly convenient optimal setting, symmetric in both local relative angles $\theta_A=\theta_B=\theta$ and detector efficiencies $\zeta_A=\zeta_B=\zeta$, with $\theta$ and $\zeta$ for $\tilde{\Delta}^{\pm}_\rho$ shown in Fig.~\ref{Delta sym}, right panel, as a function of $C$. In addition, the local rotations of the polarization vectors $\textbf{a}_1$, $\textbf{a}_2$, $\textbf{b}_1$, $\textbf{b}_2$ with respect to the settings shown in Fig.~\ref{Entangler-detector} are uniquely defined by $|\psi\rangle$.

Answering question (ii) for mixed states, we find that in contrast to pure states, not all entangled states can be detected by $W^{(2)}$. However, $\Delta^{W\pm}_\rho$ typically varies smoothly with $\rho$, i.e., mixed states sufficiently close to a detectable entangled pure state can be detected. As an example, illustrated in Fig.~\ref{Delta for different witnesses}, we consider $\rho' = \lambda \mathbb{1}/4+(1-\lambda)\rho$, with $0\leq \lambda \leq 1$, a statistical mixture of a pure state $\rho$, with concurrence $C$, and white noise. The mixed state $\rho'$ is detectable by $W^{(2)}$ if $\lambda < \Delta^{W\pm}_\rho/(\Delta^{W\pm}_\rho+|\Delta^{W\pm}_I|)$, where $\Delta^{W\pm}_I$ is the detection marginal in Eq.~\eqref{Witness condition} (possibly negative) with $\rho$ substituted by $\mathbb{1}$. We note that $\rho'$ is entangled for $\lambda < 2 C/(1 + 2 C)$, a less strict condition on $\lambda$.

Importantly, as pointed out, maximally entangled states, as the Bell states, cannot be detected by $W^{(2)}$ since for these states $2(1-\zeta_A\zeta_B)\leq \text{tr}(W^{(2)}\rho)\leq2(1+\zeta_A\zeta_B)$, which, for any $W^{(2)}$, is inside the interval of expectation values obtained from the separable states. The states are non-detectable also when non-equal weights of the two correlation measurements are used in Eq.~\eqref{Witness form}. However, by adding a third measurement, still only between two of the detector terminals and with polarization vectors in the same plane, the resulting $W^{(3)}$ can detect the maximally entangled states. For a symmetric setup, choosing a third vector along the local $z$-axis, the lowest possible detector efficiency allowing detection of the Bell states is $\zeta = 1/\sqrt{3}$, in agreement with \cite{PhysRevB.89.125404}.

In conclusion, we have shown that only two current cross correlation measurements are needed to detect entanglement of flying electron qubits. Moreover, all pure entangled states can be detected, except the maximally entangled, which require three measurements. Our findings will facilitate an experimental detection of electronic entanglement and motive further theoretical investigations on witness based entanglement quantification \cite{PhysRevA.72.022310,1367-2630-9-3-046,PhysRevLett.106.020401} and non-linear witnesses \cite{PhysRevLett.96.170502,PhysRevA.85.062327} from current cross correlations.

We acknowledge support from the Swedish Research Council. We thank F. Lefloch and C. Sch\"onenberger for discussions.

\bibliographystyle{apsrev4-1}
\bibliography{sources}

\end{document}